Initial condition of scalar perturbation in inflation

Shiro Hirai

Osaka Electro-Communication Junior College
Neyagawa, Osaka 572-8530, Japan
*E-mail*: hirai@isc.osakac.ac.jp

Abstract

A formula for the power spectrum of curvature perturbations having any initial conditions in inflation is obtained. Based on the physical conditions before inflation, the possibility exists that the initial state of scalar perturbations is not only the Bunch-Davies state, but also a more general state (a squeezed state). For example, the derived formula for the power spectrum is calculated using simple toy cosmological models. When there exists a radiation-dominated period before inflation, the behavior of the scalar perturbation is revealed not to vary greatly; however, from large scales to small scales the power spectrum of the curvature perturbations oscillates around the normal value. In addition, when inflation has a large break and the breaking time is a radiation-dominated period, a large enhancement is revealed to occur which depends on the length of the breaking time.

1. Introduction

One of the most interesting problems in cosmology is that of scalar and tensor perturbations in inflation, which has been investigated extensively. The standard equation for the power spectrum of the curvature perturbations which are produced during inflation is written as $P_R^{1/2} = (\frac{H^2}{2\pi\dot{\phi}})|_{k=aH}$, in the case of the single inflaton field, where $H$ is the Hubble parameter and $\dot{\phi}$ is the time derivative of the inflation field $\phi$.[1,2] This value is believed to be fixed at the horizon crossing in inflation and hardly changes until the perturbation reenters the Hubble horizon in the radiation-dominated period (matter-dominated period). The first and second-order corrections to this power spectrum have been investigated and were obtained in a slow-roll expansion.[3,4] A model inflation is temporarily suspended is studied and a large amplification of the curvature perturbation relative to its value at horizon crossing has been found.[5] In some cases, the enhancement of the amplitude of curvature perturbations on super horizon scales exists, and the condition for a significant effect is that the quantity $a\dot{\phi}/H$ (=$Z$) becomes sufficiently small.[6] In this study, the Hubble crossing ($\eta = \eta_*$, where $\eta$ is the conformal time), the general solution of the curvature perturbation was expressed as $\boldsymbol{R}(\eta_*) = \alpha$ $u(\eta_*) + \beta$ $v(\eta_*)$ ($u(\eta)$ is the growing mode and $v(\eta)$ is the decaying mode). At the Hubble crossing $\boldsymbol{R}(\eta_*)$ is almost cancel and after the decaying mode become negligible, then $\boldsymbol{R}(\eta)$ becomes $\alpha$ $u(\eta)$. The case $\boldsymbol{R}(\eta_*) < \boldsymbol{R}(\eta)$ arises, i.e., a large enhancement occurs as compared to its value at horizon crossing. Here, we consider the power spectrum of the curvature perturbations from another viewpoint. In a previous study [7], we describe the importance of initial condition concerning the parametric resonance. As the initial condition of the preheating scalar field, the positive-frequency solution is usually adopted in the calculation of time development of the scalar field.[8] Since inflation exists before preheating, the influence of inflation on the preheating scalar field must be taken into account. Although large squeezing on the preheating scalar field occurs during inflation, the difference in behavior of the parametric resonance between two initial conditions was shown to be not so large. Here, we consider the effect of the initial condition in inflation on the power spectrum of the curvature perturbations. In order to obtain the standard result, the action for linear scalar perturbations is described in terms of the gauge invariant potential $u$ and the field equation for $u_k$, which is the Fourier transform of $u$, is obtained. The curvature perturbation can be derived from the solution of $u_k$. Usually, in the limit $\eta \rightarrow -\infty$, the field $u_k$ approaches plane waves, i.e., the Bunch-Davies vacuum is adopted in inflation. However, we infer from the string theory/M-theory that several physical circumstances may exist before inflation, and a phase transition may be included. The importance of initial condition of inflation [9] and trans-Planckian problem of inflationary cosmology [10] are therefore discussed. We presume that there is a small possibility that the initial state of $u_k$ in inflation is the Bunch-Davies state. Herein, we assume that the initial state of $u_k$ is a more general state (a squeezed state) and investigate the difference in the power spectrum of the curvature perturbation that arises between a squeezed state and the Bunch-Davies state. A simple formula is obtained by multiplying the familiar formation by a factor which indicates the contribution of the initial condition. Using the derived formula, we estimate the effect of initial conditions using three toy cosmological models. The first model is that in which a radiation-dominated period exists before inflation. The second model is that in which the scalar matter-dominated period exists before inflation, and the third model is that in which the inflation has a long break.

In section 2, we obtain a formula for the power spectrum of curvature perturbations having any initial conditions in inflation. In section 3, using the derived formula, we estimate the effect of initial conditions using three toy cosmological models. In section 4, we discuss the results obtained in the present study.

## 2. Scalar perturbations

In this section, we obtain a formula for the power spectrum of the curvature perturbations in inflation having any initial conditions. In the present study, we apply a commonly used method.[3] For the background spectrum, we consider a spatially flat Friedman-Robertson-Walker (FRW) universe described by the metric perturbations. The line element for the background and scalar metric perturbations is generally expressed as [11]

$$ds^2 = a^2(\eta)\{(1+2A)d\eta^2 - 2\partial_i B dx^i d\eta - [(1-2\Psi)\delta_{ij} + 2\partial_i\partial_j E]dx^i dx^j\}, \qquad (1)$$

where $\eta$ is the conformal time. The density perturbation in terms of the intrinsic curvature perturbation of comoving hypersurfaces is given by

$$\boldsymbol{R} = -\Psi - \frac{H}{\dot{\phi}}\delta\phi \qquad (2)$$

where $\phi$ is the inflaton field, $\delta\phi$ is the fluctuation of the inflaton field, $H$ is the Hubble expansion parameter, and $\boldsymbol{R}$ is the curvature perturbation. Overdots represent derivatives with respect to $t$. If the gage-invariant potential $u \equiv a[\delta\phi + \frac{\dot{\phi}}{H}\Psi]$ is introduced, the action (Lagrangian) for scalar perturbations is written as [12]

$S = \int d^4 x L$

$$= \frac{1}{2} \int d\eta d^3 x \{ (\frac{\partial u}{\partial \eta})^2 - c_s^2 \ (\nabla u)^2 + \frac{Z''}{Z} u^2 \ \}, \tag{3}$$

where $c_s$ is the sound velocity, $Z = \frac{a\dot{\phi}}{H}$, and $u = -Z\boldsymbol{R}$. The field $u(\eta, \boldsymbol{x})$ is expressed using

annihilation and creation operators

$$u(\eta, x) = \frac{1}{(2\pi)^{3/2}} \int d^3 k \{ \ \boldsymbol{a_k} \ u_k(\eta) + \boldsymbol{a_{-k}}^\dagger u^*_k(\eta) \} e^{-ikx}. \tag{4}$$

The field equation of $u_k$ is derived

$$\frac{d^2 u_k}{d\eta^2} + (c_s^2 k^2 - \frac{1}{Z} \frac{d^2 Z}{d\eta^2}) \ u_k = 0. \tag{5}$$

The solution $u_k$ satisfies the normalization condition $u_k \ du^*_k / d\eta \ - u^*_k \ du_k / d\eta = i$. As an inflation we consider the power-law inflation $a(\eta) \approx (-\eta)^p \ (= t^{p/(p+1)})$. Then, equation

(5) is written as

$$\frac{d^2 u_k}{d\eta^2} + ( \ k^2 - \frac{p(p-1)}{\eta^2}) \ u_k = 0, \tag{6}$$

where in the scalar field case $c_s^2 = 1$. The solution of equation (6) is written as

$$f_k^I(\eta) = i \frac{\sqrt{\pi}}{2} e^{-ip\pi/2} (-\eta)^{1/2} \ H^{(1)}_{-p+1/2}(-k\eta) \tag{7}$$

where $H_{-p+1/2}^{(1)}$ is the Hankel function of the first kind of order $-p+\dfrac{1}{2}$. As a general initial condition, the mode function $u_k(\eta)$ is assumed to be

$$u_k(\eta)=c_1\ f_k^I\ (\eta)+c_2\ f_k^{I*}(\eta)\ , \tag{8}$$

where the coefficients $c_1$ and $c_2$ obey the relation $/c_1/^2 - /c_2/^2 = 1$. The important point is that the coefficients $c_1$ and $c_2$ do not change during inflation. In ordinary cases, the field $u_k(\eta)$ is adopted as the Bunch-Davies state, i.e. $c_1 =1$ and $c_2 =0$, because as $\eta \to -\infty$, the field $u_k(\eta)$ must approach plane waves, e.g. $e^{-ik\eta}/\sqrt{2k}$.

Next, the power spectrum is defined as [3]

$$<\boldsymbol{R}_k\ (\eta),\boldsymbol{R}_l*\ (\eta)>=\frac{2\pi^2}{k^3}\,P_R\ \delta^3(\boldsymbol{k\text{-}l}) \tag{9}$$

where $\boldsymbol{R}_k(\eta)$ is the Fourier series of the curvature perturbation $\boldsymbol{R}$. Then, the power spectrum $P_R^{1/2}$ is written as [3]

$$P_R^{1/2}=\sqrt{\frac{k^3}{2\pi^2}}\,|\frac{u_k}{Z}|. \tag{10}$$

Here, we calculate the power spectrum, using the approximation of the Hankel function. The series $H_{-p+1/2}^{(1)}(z)$ at the limit of $z \to 0$ ($z=-k\eta$) is written as

$$H_{-p+1/2}^{(1)}(z)=z^{p-1/2}\,(-\frac{i2^{1/2-p}\,sec\ p\pi}{\Gamma(1/2+p)}+\frac{i2^{-3/2-p}\,z^2\,sec\ p\pi}{\Gamma(3/2+p)}-\frac{i2^{-9/2-p}\,z^4\,sec\ p\pi}{\Gamma(5/2+p)}$$

$$+o[\,z\,]^6\,)$$

$$+ z^{-p-1/2} ( \frac{2^{-1/2+p} ( 1 + i \tan p\pi ) z}{\Gamma ( 3/2 - p )} - \frac{i 2^{-5/2+p} ( -i + \tan p\pi ) z^3}{\Gamma ( 5/2 - p )} - \frac{2^{-11/2+p} ( 1 + i \tan p\pi ) z^5}{\Gamma ( 7/2 - p )}$$

$$+ o[ z ]^7 ),$$

$$(11)$$

where $\Gamma ( -p + 1/2 )$ expresses the Gamma function. In the present study, the term $z^{p-1/2}$ $(=(-k\eta )^{p-1/2})$ is of leading order. The power spectrum of the leading and next leading correction of $|-k\eta|$ in the case of squeezed initial states can be written as

$$P_R^{1/2} = ( 2^{-p} ( -p )^p \frac{\Gamma ( -p + 1/2 )}{\Gamma ( 3/2 )} \frac{1}{m_P^2} \frac{H^2}{|H'|} )|_{k=aH} ( 1 - \frac{( -k\eta )^2}{2( 1 + 2p )} )$$

$$\times | c_1 e^{-ip\pi /2} + c_2 e^{ip\pi /2} |.$$

$$(12)$$

We define the quantity $C(k)$ as

$$C(k) = c_1 e^{-ip\pi /2} + c_2 e^{ip\pi /2}.$$

$$(13)$$

If $|C(k)| = 1$, the leading term of $-k\eta$ of equation (12) can be written as $P_R^{1/2} = ( \frac{H^2}{2\pi\dot\phi} )|_{k=aH}$ [13]. However, equation (12) implies that if, under some physical circumstances, the Bunch-Davies state is not adopted as the initial condition of the field $u_k$, the possibility exists that $|C(k)| \neq 1$. Then, the next problem is whether such a situation actually exists. Here, we assume a number of pre-inflation toy models. For example, the situation that the pre-inflation is a radiation-dominated period is quite probable. We know that a state of $u_k$ becomes a squeezed state at the initial time of inflation, i.e., the state of $u_k$ is not the Bunch-Davies state. Calculation of $|C(k)|$ in this case yields interesting results, even though the effect of $|C(k)|$ on the value of $|C(k)|$ is small. In following sections, we shall calculate the value of $|C(k)|$ using a number of pre-inflation models.

## 3. Models of pre-inflation

In the previous section, we showed that the difference in initial states of the scalar perturbations in inflation influences the asymptotic form of the power spectrum. We now investigate the type of pre-inflation period conditions required in order to bring about a squeezed initial state in inflation. Here, we consider simple toy models of a pre-inflation, i.e., the pre-inflation period is a radiation-dominated period and a scalar-matter-dominated period. Furthermore, we consider the inflation model having a break in which the break period is a radiation–dominated period (a double inflation). Before presenting examples, a simple cosmological model is assumed as

$$a^{I1} = a_0(-\eta - \eta_i)^q,$$

$$a^{R1} = a_1(-\eta - \eta_j)^r,$$

$$a^{I2} = a_2(-\eta)^p,$$

$$a^{R2} = a_3(\eta - \eta_m),$$

$$a^{M} = a_4(\eta - \eta_n)^2, \tag{14}$$

where,

$$\eta_i = (\frac{q}{r} - 1)\eta_1 + \frac{q}{r}\eta_j \quad , \quad \eta_j = (\frac{r}{p} - 1)\eta_2 \quad , \quad \eta_m = \eta_3(1 - \frac{1}{p}) \quad , \quad \eta_n = -\eta_4 + 2\eta_3(1 - \frac{1}{p}) \quad ,$$

$$a_0 = \frac{r}{q}\frac{(-\eta_1 - \eta_j)^{r-1}}{(-\eta_1 - \eta_i)^{q-1}}a_1, a_1 = (\frac{p}{r})^r(-\eta_2)^{p-r}a_2, a_3 = -p(-\eta_3)^{p-1}a_2, a_4 = \frac{a_3}{2(\eta_4 - \eta_n)}.$$

The scale factor $a^{I2}$ represents the ordinary inflation. If $p$=-1, the inflation is de-Sitter inflation, and if $p$<-1, the inflation is a power-law inflation. We assume that inflation begins at $\eta = \eta_2$ and ends at $\eta = \eta_3$. The radiation-dominated period (in which the scale factor is $a^{R2}$) follows, and the matter-dominated period (in which the scale factor is $a^{M}$) begins at $\eta = \eta_4$. Finally, $\eta_5$ is the current time. We now consider the period before inflation. In pre-inflation, if the case is $r$=1, the scale factor $a^{R1}$ indicates that the radiation–dominated period occurs, and if the case is $r$=2, the scale factor $a^{R1}$ indicates that the scalar-matter-dominated period occurs. If we considered the inflation having a break, the scale factor $a^{I1}$ is needed. In this case, we consider the power-law inflation. If a sufficient long inflation exists, the physics of pre-inflation only slightly affects the cosmological perturbations. Using simple toy models, we investigate the validity of this statement. Here, the period of the second inflation is assumed to be sufficiently long, i.e., when we plot the graph of $|C(k)|$, we choose the value of $\eta_2$ as the time when perturbations of the current Hubble horizon size exceeds the Hubble radius in inflation. For example, we show the values of $\eta_5$, $\eta_4$, $\eta_3$, $\eta_2$, $\eta_1$, i.e., $\eta_5$=1.25×10$^{18}$, $\eta_4$=1.66×10$^{16}$, $\eta_3$ =-6.63×10$^{-9}$, and the value of $\eta_2$ depends on the value of $p$, if $p$=-$\frac{10}{9}$ ($a$=$t^{10}$), then

$\eta_2$ =-7.03×10$^{17}$, and if $p$=-$\frac{100}{99}$ ($a$= $t^{100}$), then $\eta_2$ =-6.39×10$^{17}$, and we set $\eta_1$=$s\eta_2$,

where $s$ is a parameter of the length of the breaking time between two inflations.

## 3.1 Case of radiation-dominated period before inflation

We consider the case that a radiation-dominated period before inflation exists. In this case,

the scale factor becomes $a^{R1}$=$a_1$(-$\eta$ -$\eta_j$), i.e., $r$=1. In the radiation-dominated period the filed equation $u_k$ can be written as (5). In this case $Z$ is written as $Z$= $a^{R1}$ (2($\boldsymbol{H}^2$ -

$\boldsymbol{H'}$ )/3)$^{1/2}$ /($c_s$) [12,14] where $\boldsymbol{H}$ =( $a^{R1}$ )$'$ / $a^{R1}$, and the prime denote the derivative with

respect to conformal time $\eta$ . We can fix the value of $c_s^2$ as 1/3. The solution of equation

(5) is thus given as [14]

$$f_k^R (\eta)=\frac{3^{1/4}}{\sqrt{2k}} e^{-ik(\eta+\eta_j)/\sqrt{3}} . \tag{15}$$

Here, for simplicity we assume that the mode function of the radiation-dominated period

can be written as equation (15). The solution of equation (6) in inflation becomes equation

(7). Then, the general mode function in inflation can be written as

$$u_k^I(\eta)= c_1 f_k^I(\eta)+c_2 f_k^{I*}(\eta). \tag{16}$$

In order to fix the coefficients $c_1$ and $c_2$, we use the matching condition that the mode function and first $\eta$ –derivative of the mode function are continuous at the transition time $\eta = \eta_2$. Then, the coefficients $c_1$ and $c_2$ are calculated as

$$c_1 = \frac{\sqrt{\pi}}{2 3^{3/4} \sqrt{2z}} e^{i(p\pi/2 + 2z/\sqrt{3}p)} ((3-3p-i\sqrt{3}z) H^{(2)}_{-p+1/2}(z) - 3z H^{(2)}_{-p+3/2}(z)), \tag{17}$$

$$c_2 = \frac{\sqrt{\pi}}{2 3^{3/4} \sqrt{2z}} e^{i(-p\pi/2 + 2z/\sqrt{3}p)} ((3-3p-i\sqrt{3}z) H^{(1)}_{-p+1/2}(z) - 3z H^{(1)}_{-p+3/2}(z)), \tag{18}$$

where $z = -k\eta_2$. The quantity $C(k)$ is derived from equation (13) as

$$C(k) =$$
$$-\frac{\sqrt{\pi}}{2 3^{3/4} \sqrt{2z}} e^{2iz/\sqrt{3}p} \{(-3+3p+i\sqrt{3}z)(H^{(1)}_{-p+1/2}(z) + H^{(2)}_{-p+1/2}(z)) + 3z(H^{(1)}_{-p+3/2}(z) + H^{(2)}_{-p+3/2}(z))\}$$

$$\tag{19}$$

The quantity $|C(k)|$, which shows that a contribution of pre-inflation ($k$-dependence) is plotted as a function of $z$ ($=-k\eta_2$) in figure 1, where the case $p$=-10/9 is considered. From figure 1, we find that when $z<1$, $C(k)$ becomes zero, and when $z$=1.11 ($k$=1.58×10$^{-18}$), the value of $|C(k)|$ is 0.737. Finally, when $z>1$, $|C(k)|$ oscillates around 1. Since we are interested in the super-large scale, i.e. $z \ll 1$, we expand the Hankel function around $z$=0 (equation (9)). In the case of $z \to 0$, the quantity $C(k)$ can be obtained in simple form as

$$C(k) \cong \frac{2^{-1+p} \sqrt{\pi}}{3^{3/4} \Gamma(\frac{3}{2} - p)} e^{2iz/\sqrt{3}p} z^{-p} (3-3p-i\sqrt{3}z). \tag{20}$$

If we consider the case p=-10/9, the coefficients $|C(k)|$ are proportional to $z^{10/9}$, when z $\rightarrow 0$ ($k \rightarrow 0$), and $|C(k)|$ becomes zero. Next, we consider the case of z$\rightarrow \infty$ (small-scale cases), for which the quantity $|C(k)|$ is approximately

$$|C(k)| \cong \{ \frac{1}{\sqrt{3}}(2- cos(\frac{\pi}{9} - 2z))\}^{1/2} . \tag{21}$$

From equation (21), the leading contribution of $|C(k)|$ does not depend on p, and oscillates around 1. The numerical value is obtained as $0.760 \le |C(k)| \le 1.32$ from equation (21).

Here, we assume that inflation starts at $\eta_2$ ($=- z / k$) and $\eta_2$ is the time when the perturbations of the current Hubble horizon size exceeds the Hubble radius in inflation. However, if longer inflation exists, from the super horizon scales to the small scales, the quantity $|C(k)|$ behaves as in equation (21), i.e., $|C(k)|$ oscillates around 1. Although the results from existence of the radiation-dominated period before inflation are not altogether surprising, a very long inflationary period can not remove the vibration of $|C(k)|$ around 1.

### 3.2 Case of the scalar-matter-dominated period before inflation

We consider the case of the existence of a scalar-matter-dominated period before inflation in which the scalar-matter is the inflaton field $\phi$. In this case, the scale factor becomes $a^{R1}=a_1(-\eta -\eta_j)^2$, i.e., r=2. The solution of equation (6) is given as

$$f_k^S (\eta)=\frac{1}{\sqrt{2k}}(1- \frac{i}{k(\eta +\eta_j)})e^{-ik(\eta+\eta_j)} . \tag{22}$$

Here, for simplicity, we assume that the mode function of the scalar-dominated period can be written as $f_k^s(\eta)$. The solution of equation (6) in inflation is equation (7). Then, the general mode function in inflation can be written as (8). In order to fix the coefficients $c_1$ and $c_2$, we assume that the mode function and first $\eta$–derivative of the mode function are continuous at $\eta = \eta_2$. Then, the coefficients $c_1$ and $c_2$ are calculated as

$$c_1 = \frac{-i\sqrt{\pi}}{8\sqrt{2z^3}} e^{i(p\pi/2+2z/p)} (( p^2 +4z(i+z)-2p(1+iz))\, H_{-p+1/2}^{(2)}(z) +2(p-2iz)z\ H_{-p+3/2}^{(2)}(z)), \quad (23)$$

$$c_2 = \frac{-i\sqrt{\pi}}{8\sqrt{2z^3}} e^{i(-p\pi/2+2z/p)} (( p^2 +4z(i+z)-2p(1+iz))\, H_{-p+1/2}^{(1)}(z) +2(p-2iz)z\ H_{-p+3/2}^{(1)}(z)). \quad (24)$$

The quantity $C(k)$ is derived from equation (13) as

$$C(k) = \frac{-i\sqrt{\pi}}{8\sqrt{2z^3}} e^{2iz/p} \{( p^2 +4z(i+z)-2p\,(1+iz))\, (\, H_{-p+1/2}^{(1)}(z)+\ H_{-p+1/2}^{(2)}(z))$$
$$+ 2(p-2iz)z\, (\, H_{-p+3/2}^{(1)}(z)+H_{-p+3/2}^{(2)}(z))\, \}. \quad (25)$$

The quantity $|C(k)|$, which shows the contribution of pre-inflation ($k$-dependence), is plotted in figure 2 as a function of $z\, (=-k\eta_2)$ for the case of $p$=-10/9. From figure 2 we find that when $z<1$, $C(k)$ becomes zero, and that when $z$=1.11 ($k$=1.58×10$^{-18}$), the value of $|C(k)|$ is 0.6239. In the case $z\gg1$, we find that $C(k)\cong 1$, i.e., enhancement does not occur. Since we are interested in the super-large scale, in which $z\ll1$, we expand the Hankel function around $z$=0. Using equation (11), $C(k)$ can be obtained in simple form as

$$C(k) \cong -\frac{2^{-3+p} i\sqrt{\pi}\, e^{2iz/p} ( p^2 -2p(1+iz)+4z(i+z))z^{-1-p}}{\Gamma(\frac{3}{2}-p)}. \quad (26)$$

If we consider the case in which $p$=-10/9, the coefficients $|C(k)|$ are proportional to $z^{1/9}$, when $z \to 0$ ($k \to 0$), and so $|C(k)|$ becomes zero. Next, we consider the case in which $z \to \infty$ (small-scales cases). Here, $|C(k)|$ is obtained as

$$|C(k)| \cong \sqrt{1 - \frac{19}{9z} sin(\frac{\pi}{9} - 2z)} \; . \tag{27}$$

In this case, $|C(k)| \cong 1$, which is different from the case of the radiation-dominated period before inflation.

## 3.3  Case of inflation having a long break

We now consider the case of an inflation having a long break. The first inflation ends at $\eta = \eta_1 = s\,\eta_2$ and the radiation-dominated period, in which $r$=1, follows. The second inflation occurs at $\eta = \eta_2$. In the radiation-dominated period, the value of $s$ is restricted to $1 < s < (p-1)/p$. We assume that the start time of the second inflation $\eta_2$ is the time when the perturbation of the current Hubble horizon size exceeds the Hubble radius in inflation. It is believed that if sufficiently long inflation exists, the physics of the pre-inflation will barely affect the cosmological perturbations. In the first inflation, the solution of equation (6) is given as $u_k^{I1}(\eta) = f_k(\eta + \eta_i)$, provided that the coefficient $p$ is replaced by $q$. For simplify, we assume that the mode function of first inflation is written as $u_k^{I1}(\eta)$, i.e., the Bunch-Davies state is adopted. Using equation (15) in the first radiation-dominated period, the mode function is written as

$$u_k^{R1}(\eta) = c_3 \, f_k^R(\eta) + c_4 \, f_k^{R*}(\eta).$$  (28)

Using the matching condition the coefficients $c_3$ and $c_4$ are calculated such as

$$c_3 = -\frac{\sqrt{\pi}}{3^{3/4}2\sqrt{2pq(1+p(-1+s))z}} \, e^{-i\pi q/2} \, e^{-i(1+p(-1+s))z/\sqrt{3}p}$$

$$(3q(-1+p-sp)z \, H_{-q+3/2}^{(1)}(cz) + (i\sqrt{3}\,qz + p(3-3q-i\sqrt{3}\,qz + i\sqrt{3}\,sqz)) \, H_{-q+1/2}^{(1)}(cz)),$$  (29)

$$c_4 = -\frac{\sqrt{\pi}}{3^{3/4}2\sqrt{2pq(1+p(-1+s))z}} \, e^{-i\pi q/2} \, e^{i(1+p(-1+s))z/\sqrt{3}p}$$

$$(3q(1-p+sp)z \, H_{-q+3/2}^{(1)}(cz) + (i\sqrt{3}\,qz + p(-3+3q-i\sqrt{3}\,qz + i\sqrt{3}\,sqz)) \, H_{-q+1/2}^{(1)}(cz))$$  (30)

where $c = -((q-2)s + q(1-p)/q)$. In the second inflation, the solution of (6) is written as (8). Using the matching condition the coefficients $c_1$ and $c_2$ are calculated such as

$$c_1 = -\frac{\sqrt{\pi}}{12\sqrt{3}\sqrt{pq(1+p(-1+s))z}} \, e^{i(p-q)\pi/2 + i(-1+s)z/\sqrt{3}}$$

$$((-3+3p-i\sqrt{3}\,z) \, H_{-p+1/2}^{(2)}(z) + 3z \, H_{-p+3/2}^{(2)}(z))$$

$$\times (-3q(1+p(-1+s))z \, H_{-q+3/2}^{(1)}(cz) + (-i\sqrt{3}\,qz + p(3-3q+i\sqrt{3}\,qz - i\sqrt{3}\,qsz)) \, H_{-q+1/2}^{(1)}(cz))$$

$$+ e^{-2i(-1+s)z/\sqrt{3}} ((-3+3p+i\sqrt{3}\,z) \, H_{-p+1/2}^{(2)}(z) + 3z \, H_{-p+3/2}^{(2)}(z)) \, (3q(1+p(-1+s))z \, H_{-q+3/2}^{(1)}(cz)$$

$$+ (-i\sqrt{3}\,qz + p(-3+3q+i\sqrt{3}\,qz - i\sqrt{3}\,qsz)) \, H_{-q+1/2}^{(1)}(cz))\},$$  (31)

$$c_2 = \frac{i\sqrt{\pi}}{12\sqrt{3}\,pq(1+p(-1+s\,))z}\; e^{-i(\,p+q\,)\pi/2+i(-1+s\,)z/\sqrt{3}}$$

$$((-3+3p-i\sqrt{3}\,z)\; H^{(1)}_{-p+1/2}(z)+ 3z\,H^{(1)}_{-p+3/2}(z))$$

$$\times\; (-3q(1+p(-1+s))z\,H^{(1)}_{-q+3/2}(cz) + (-i\sqrt{3}\;qz+p(3-3q+i\sqrt{3}\;qz-i\sqrt{3}\;qsz))\; H^{(1)}_{-q+1/2}(cz))$$

$$-\; e^{-2i(-1+s\,)z/\sqrt{3}}\,((-3+3p+i\sqrt{3}\,z)\; H^{(1)}_{-p+1/2}(z)+3z\,H^{(1)}_{-p+3/2}(z))$$

$$\times\; (3q(1+p(-1+s))z\,H^{(1)}_{-q+3/2}(cz) + (-i\sqrt{3}\;qz+p(-3+3q+i\sqrt{3}\;qz-i\sqrt{3}\;qsz))\; H^{(1)}_{-q+1/2}(cz))\}. \quad (32)$$

The quantity $C(k)$ is derived from equation (13) as

$$C(k) = -\frac{\sqrt{\pi}}{12\sqrt{3}\,pq(1+p(-1+s\,))z}\; e^{-i\pi q/2}\; e^{i(-1+s\,)z/\sqrt{3}}$$

$$\{(3z(H^{(1)}_{-p+3/2}(z)+H^{(2)}_{-p+3/2}(z))+(-3+3p-i\sqrt{3}\,z)\;(H^{(1)}_{-p+1/2}(z)+\;H^{(2)}_{-p+1/2}(z)))$$

$$\times\; (3q(-1+p(1-s))z\,H^{(1)}_{-q+3/2}(cz)+(-i\sqrt{3}\;qz+p(3+q(-3-i\sqrt{3}\;(-1+s)z)))\; H^{(1)}_{-q+1/2}(cz))$$

$$+\; e^{2i(1-s\,)z/\sqrt{3}}\,(3z(H^{(1)}_{-p+3/2}(z)+H^{(2)}_{-p+3/2}(z))+(\;(-3+3p+i\sqrt{3}\,z\,)(H^{(1)}_{-p+1/2}(z)+\;H^{(2)}_{-p+1/2}(z))$$

$$\times\; (3q(1+p(-1+s))z\,H^{(1)}_{-q+3/2}(cz)+(-i\sqrt{3}\;qz+p(-3+q(3-i\sqrt{3}\;(-1+s)z))))\; H^{(1)}_{-q+1/2}(cz)\}. \quad (33)$$

For super-large scale, in which $z \ll 1$, using equation (11), the quantity $C(k)$ can be obtained simple form such as

$$|C(k)| \cong \frac{2^{p-q}(q/p)^q(1+p(-1+s\,))^{-1+q}\,\Gamma(1/2-q)}{\Gamma(1/2-p\,)}\; z^{-p+q}\;. \quad (34)$$

For $p=q$, $|C(k)|$ is simplified for $z \to 0$ as

$$|C(k)| \cong (1 + p(s-1))^{p-1}.$$  (35)

Next, we consider the case of $z \to \infty$ (small-scale cases) for which $|C(k)|$ is obtained as

$$|C(k)|^2 \cong \{ \frac{4}{3} + \frac{1}{12} (-4\cos(\frac{2(-1+s)z}{\sqrt{3}}) + 8\cos(p\pi + 2z)$$

$$+ 2\sqrt{3} \cos((3p\pi - 3q\pi + 4\sqrt{3}(-1+s)z)/6) + \cos((3p\pi - 3q\pi - 4\sqrt{3}(-1+s)z)/6)$$

$$+ (-4 + 2\sqrt{3})\cos(p\pi + 2(3 + \sqrt{3} - \sqrt{3}s)z/3) + (-4 - 2\sqrt{3})\cos(p\pi + 2(3 - \sqrt{3} + \sqrt{3}s)z/3)$$

$$+ (-3 - 2\sqrt{3})\cos((9p\pi - 3q\pi + 4(3 + \sqrt{3}(-1+s)z)/6)$$

$$+ (-3 + 2\sqrt{3})\cos((9p\pi - 3q\pi + 4(3 + \sqrt{3}(1-s)z)/6)$$

$$+ (3 - 2\sqrt{3})\cos((3p\pi + 3q\pi + 4(3 + \sqrt{3}(1-s)z)/6)$$

$$+ (3 + 2\sqrt{3})\cos((3p\pi + 3q\pi + 4(3 - \sqrt{3}(1-s)z)/6).$$  (36)

We now present the results derived in this section 3.3. First, in order to show the overall results, we plot the quantity $|C(k)|$ as a function of $z$ ($= -k\eta_2$) in figure 3, in which we fixed $p = q = -10/9$ and $s = 1.85$. At $z = 1.11$, where the mode $k$ is the current Hubble horizon size, and the value of $|C(k)|$ is 331, considerable enhancement occurs. On the other hand, at $z \geq 20$, the value of $|C(k)|$ is of order 1. When $1 \leq z \leq 20$ the value of $|C(k)|$ oscillates and becomes smaller. As in this figure, the length of the second inflation is chosen as 60 *e-holds* (the precise value is 66.7 *e-holds*). The range of $z$, $1 \leq z \leq 20$, indicates the range of the mode $k$, $1.58 \times 10^{-18} \leq k \leq 2.8 \times 10^{-17}$ for large scales. Therefore, physically important effects may occur. However, this enhancement does not occur if the length of the second inflation is sufficiently long.

We now consider the case of $z \to 0$. Equation (35) shows that if the length $s$ of the radiation-dominated period approaches $(p-1)/p$, $|C(k)|$ becomes a large constant value. For example, when s=1.85, 1.89, and 1.899, $|C(k)|$ are calculated for the case of $p=q=-10/9$ as 447, $1.34 \times 10^3$ and $1.72 \times 10^6$, respectively. This property does not change by with the value of $p$.

We now consider the case of $z \to \infty$. From equation (36), the quantity $|C(k)|$ oscillates as $0.6 \leq |C(k)| \leq 1.7$ from large scales to small scales. This property depends only slightly on $p(=q)$ and $s$.

Finally, we consider the case of the transition region between $z \to 0$ and $z \to \infty$. The quantity $|C(k)|$ is large when $z \to 0$, and becomes of order 1 when $z \to \infty$. This transition range depends on the length of the radiation-dominated period $s$. Furthermore, we can infer that the modes $k$ (z=-$k\eta_2$) of the transition range are the modes $k$ that exceed the Hubble horizon at the first inflation, enter the Hubble horizon at the radiation–dominated period (the period between the two inflations) and again exceed the Hubble horizon at the second inflation. This is another difference from Leach's condition using $Z$.[6]

## 4. Discussion and summary

We consider the possibility that the initial state of scalar perturbation is not only the Bunch-Davies state, but also a more general state (a squeezed state) based on the physical conditions before inflation. Using a squeeze initial state, we calculated the power spectrum of the curvature perturbation. The derived formula is a commonly used formula multiplied by $|C(k)|$. (see equation (12)). Using the derived formula, we estimate the effect of the initial conditions using three toy cosmological models. The first model is that in which the radiation-dominated period before inflation exists. We consider this model to be plausible. The derived results are described in the following. If inflation is of finite length, in the super-large scales, the power spectrum of the curvature perturbation becomes zero. The effect of the radiation-dominated period before inflation is that $|C(k)|$ oscillate around 1 from large scales to small scales. Next, we consider the case in which the scalar-matter-dominated period before inflation exists. The result for this case is similar to that for the radiation-dominated period, but from large scales to small scales, $|C(k)|$ becomes 1. Finally, we consider the case of inflation having a long break. Large enhancement occurs in the case of super large scales and from large scales to small scales $|C(k)|$ oscillates around 1. The enhancement depends on the length of the break. Furthermore, the length of the transition range between the large enhancement and no enhancement appears to depend on the length of the break. Similar enhancement was derived in [5,6].

Here, we consider only the case of scalar perturbation, and will investigate tensor perturbation in a future publication. In addition, in a future study, rather than using simple toy cosmological models, we will calculate our formula using more realistic models.

Acknowledgment

The author would like thank the staff of Osaka Electro-Communication Junior College for their invaluable discussions.

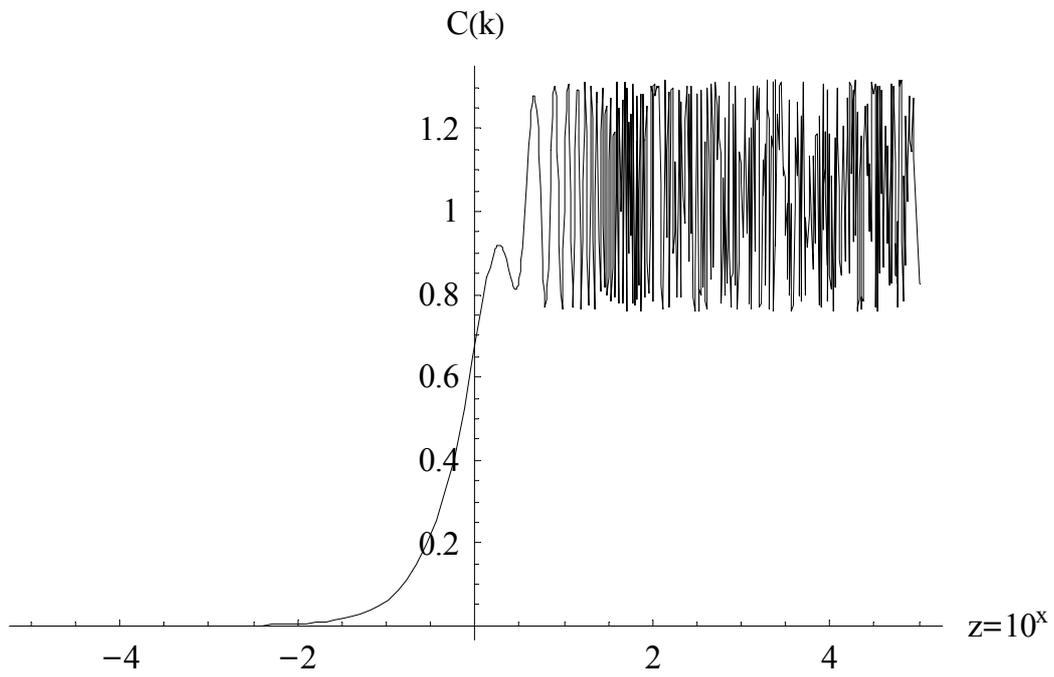

Figure 1. Enhancement factor $|C(k)|$ as the function $z(=-k\eta_2)$ for the range $10^{-5} \leq z \leq 10^5$ in the case of the radiation-dominated period before inflation, where $p = -\dfrac{10}{9}$ .

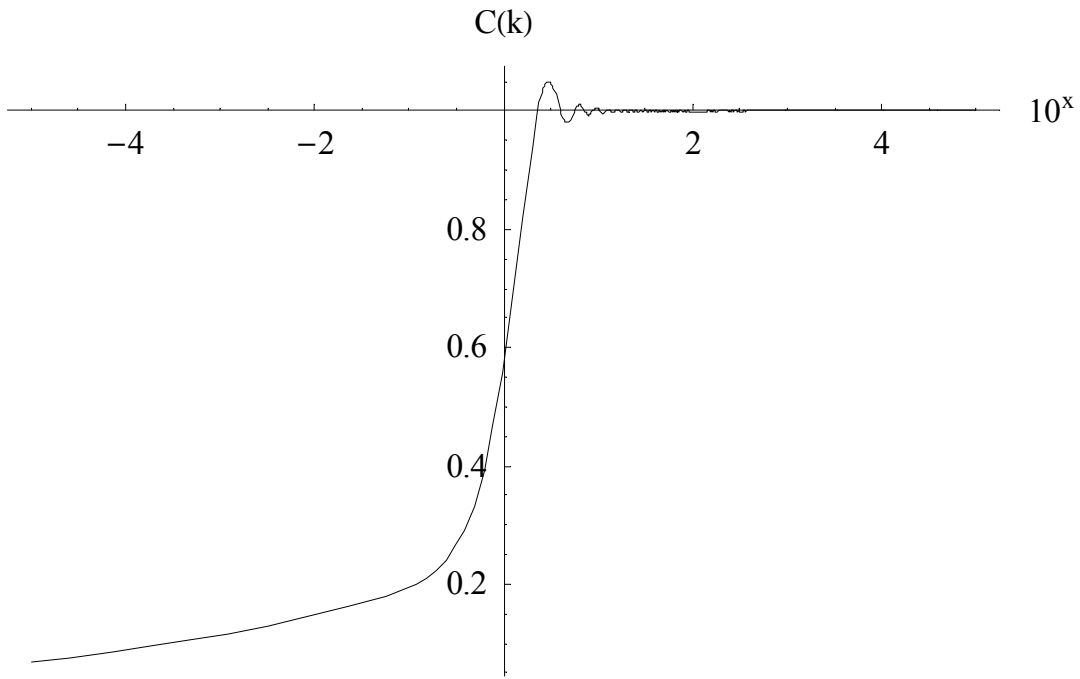

Figure 2. Enhancement factor $|C(k)|$ as the function $z(=-k\eta_2)$ for the range $10^{-5} \leq z \leq 10^5$ in the case of the scalar-matter-dominated period before inflation, where $p=-\dfrac{10}{9}$.

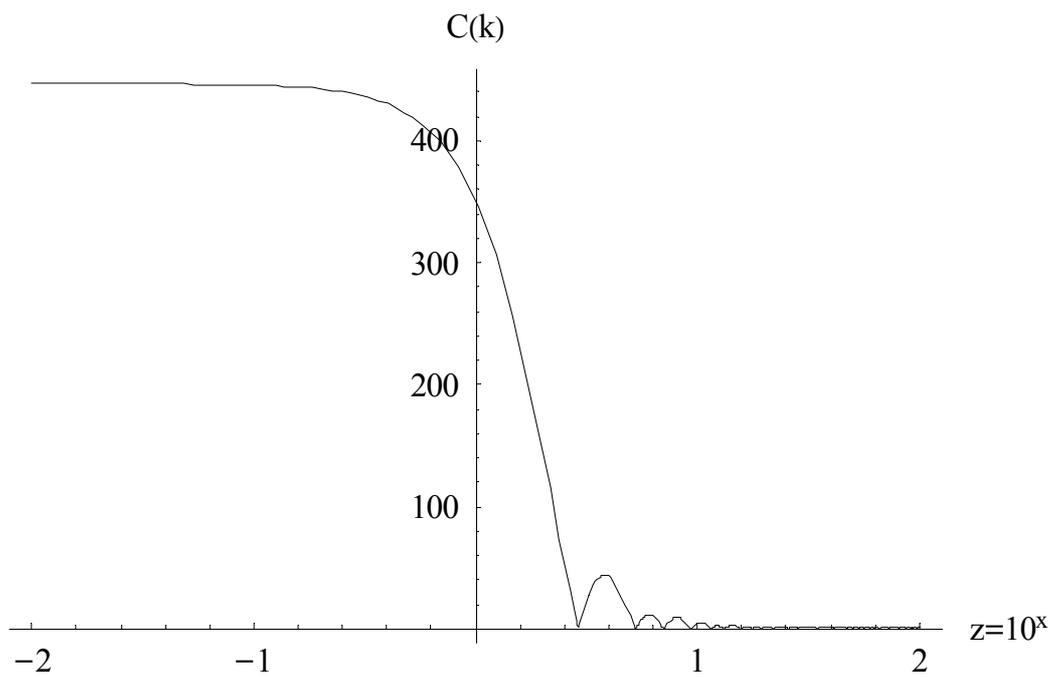

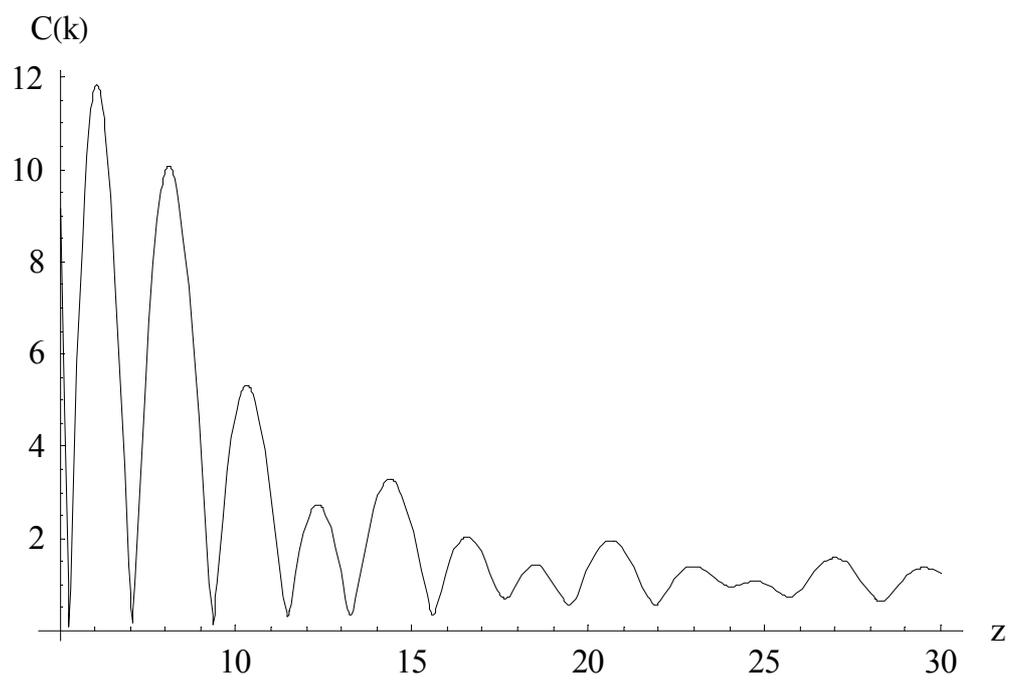

Figure 3. Enhancement factor $|C(k)|$ as the function $z(=-k\eta_2)$ for the ranges $10^{-2} \leq z \leq 10^2$ and $5 \leq z \leq 30$ in the case of inflation having a long break, where $p=q=-\dfrac{10}{9}$ and $s=1.85$.